\documentstyle[11pt]{article}

\begin{document}

\title{\vspace*{-1cm} {\bf Parity-dependent squeezing of light}}
\author{{\large 
C Brif,$^{1}$\thanks{E-mail: costya@physics.technion.ac.il}  
\ A Mann,$^{1}$\thanks{E-mail: ady@physics.technion.ac.il} 
\ and A Vourdas$\,{}^{2}$\thanks{E-mail: ee21@liverpool.ac.uk} } 
\vspace*{0.2cm} \\ 
{\em ${}^{1}$Department of Physics, Technion -- Israel 
Institute of Technology, Haifa 32000, Israel} \\
{\em ${}^{2}$Department of Electrical Engineering and Electronics,
University of Liverpool,} \\
{\em Brownlow Hill, Liverpool L69 3BX, United Kingdom}}
\date{to appear in {\em Journal of Physics A: Mathematical and 
General}}
   \maketitle

\vspace*{0.3cm}

	\begin{abstract}
\noindent
A parity-dependent squeezing operator is introduced which imposes
different SU(1,1) rotations on the even and odd subspaces of the
harmonic oscillator Hilbert space. This operator is used to define
parity-dependent squeezed states which exhibit highly nonclassical
properties such as strong antibunching, quadrature squeezing, strong
oscillations in the photon-number distribution, etc. In contrast to
the usual squeezed states whose $Q$ and Wigner functions are 
simply Gaussians, the parity-dependent squeezed states have much more
complicated $Q$ and Wigner functions that exhibit an interesting
interference in phase space. The generation of these states
by parity-dependent quadratic Hamiltonians is also discussed.
	\end{abstract}

\section{Introduction}
\setcounter{equation}{0}

\noindent
Squeezed states have been studied extensively in the last few years 
%\cite{SS1,SS2,SS3,SMD}
[1--4]. They exhibit nonclassical behavior such 
as oscillations in the photon-number distribution \cite{oscil}, 
sub-Poissonian photon statistics (antibunching),
reduction of quantum fluctuations in either of the field quadratures
(quadrature squeezing), etc. Squeezing of the two-mode light field 
was studied in \cite{SS2}, and squeezing criteria for multi-mode systems
were introduced in \cite{SMD}. It was pointed out \cite{Vou} that the 
usual type of two-mode squeezing defined in \cite{SS2} is based on 
reducible representations of the SU(1,1) group. A more general kind of 
two-mode squeezing was considered \cite{Vou} where different SU(1,1) 
rotations are imposed on each irreducible sector. It was shown 
\cite{Vou} that the two-mode squeezed states produced by these 
generalized squeezing transformations have interesting properties. 

In the present paper we study a similar generalization of squeezing 
for the single-mode light field. The ordinary single-mode squeezed 
states are produced by unitary transformations belonging to a 
reducible representation of SU(1,1). Such reducible representations 
contain two irreducible components: the first is the representation 
that acts on the ``even Fock subspace'' (i.e., the subspace spanned 
by the even number eigenstates); and the second is the 
representation that acts on the ``odd Fock subspace'' (i.e., the 
subspace spanned by the odd number eigenstates). In terms of the
Bargmann index $k$ that labels unitary irreducible representations 
of SU(1,1) \cite{SU11}, the even Fock subspace corresponds to the 
representation with $k=1/4$, while the odd Fock subspace 
corresponds to the case $k=3/4$. We impose different SU(1,1) 
rotations on the two irreducible sectors and thus introduce states
whose even and odd components are squeezed with different squeezing
parameters. We refer to these states as the parity-dependent 
squeezed states.

The parity-dependent squeezed states exhibit highly nonclassical 
behavior. They are characterized by more squeezing parameters than 
the ordinary squeezed states, and we find regions of these 
parameters where there are strong oscillations in the photon-number 
distribution, strong antibunching, quadrature squeezing, etc. We 
demonstrate that these states can be more strongly antibunched than 
the ordinary squeezed states. A further interesting feature of the 
parity-dependent squeezed states is that their $Q$ and Wigner 
functions are not necessarily Gaussians. The examples that we 
consider show a very strong interference in phase space.

\section{Parity-dependent squeezed states}
\setcounter{equation}{0}

\subsection{Definitions and basic properties}

\noindent
We consider the harmonic oscillator Hilbert space ${\cal H}$ and
express it as the direct sum
\begin{equation}
{\cal H} = {\cal H}_{0} \oplus {\cal H}_{1}   \label{2.1}
\end{equation}
where ${\cal H}_{0}$ and ${\cal H}_{1}$ are the subspaces spanned 
by the even and odd number eigenstates, respectively:
\begin{equation}
\begin{array}{l}\vspace{0.2cm}
{\cal H}_{0} = \{ |2n\rangle ; n=0,1,2,\ldots \}  \\
{\cal H}_{1} = \{ |2n+1\rangle ; n=0,1,2,\ldots \} .
\end{array}    \label{2.2}
\end{equation}
The projection operators onto these Hilbert spaces are given by
\begin{equation}
\begin{array}{l}\vspace{0.2cm}
\Pi_{0} = \displaystyle{ \sum_{n=0}^{\infty} }
|2n \rangle\langle 2n|   \\
\Pi_{1} = \displaystyle{ \sum_{n=0}^{\infty} }
|2n+1 \rangle\langle 2n+1| .
\end{array}    \label{2.3}
\end{equation}
They have the usual properties of projection operators:
\begin{equation}
\begin{array}{l}\vspace{0.2cm}
\Pi_{0} + \Pi_{1} = 1  \\
\Pi_{i}\Pi_{j} = \Pi_{i} \delta_{ij} \;\;\;\;\;\;\;\;\;\; 
i,j = 0,1 .
\end{array}    \label{2.4}
\end{equation}
The parity operator is given by
\begin{equation}
P = \Pi_{0} - \Pi_{1} = \exp(i\pi a^{\dagger}a)   \label{2.5}
\end{equation}
where $a$ and $a^{\dagger}$ are the boson annihilation and creation
operators. The properties of the parity operator are
\begin{equation}
\begin{array}{l}\vspace{0.2cm}
P^{2} = 1 \\   \vspace{0.2cm}
Pa = -aP \\
Pa^{\dagger} = -a^{\dagger}P .
\end{array}    \label{2.6}
\end{equation}
If $|\Psi\rangle$ is an arbitrary state in the Hilbert space 
${\cal H}$ with the position-representation wave function 
$\Psi(x) = \langle x|\Psi\rangle$, then the action of the parity
operator is the inversion:
\begin{equation}
P \Psi(x) = \langle x|P|\Psi\rangle = \Psi(-x) . \label{2.7}
\end{equation}
The same property also holds for the momentum representation.

The ordinary single-mode squeezing operator is defined as \cite{SS1}
\begin{equation}
\begin{array}{c}\vspace{0.2cm}
S(\xi,\lambda) = \exp\left(\xi K_{+} - \xi^{\ast} K_{-} \right)
\exp(2i\lambda K_{0})  \\
\xi = -r \exp(-i\theta) .
\end{array}     \label{2.8}   
\end{equation}
The parameters $r$, $\theta$, and $\lambda$ are real. The operators
\begin{equation}
K_{0} = \frac{1}{4} (aa^{\dagger}+a^{\dagger}a)  \;\;\;\;\;\;\;\;\;\;
K_{+} = \frac{a^{\dagger 2}}{2}  \;\;\;\;\;\;\;\;\;\;
K_{-} = \frac{a^{2}}{2}       \label{2.9}
\end{equation}
form the single-mode bosonic realization of the SU(1,1) Lie algebra:
\begin{equation}
[K_{0},K_{\pm}] = \pm K_{\pm}   \;\;\;\;\;\;\;\;\;\; 
[K_{-},K_{+}] = 2K_{0} .     \label{2.10}
\end{equation}
The Casimir operator is
\begin{equation}
K^{2} = K_{0}^{2}-\frac{1}{2}(K_{+}K_{-}+K_{-}K_{+}) 
= k(k-1) = -\frac{3}{16}   \label{2.11}
\end{equation}
where the Bargmann index $k$ labels the irreducible representations
of SU(1,1) \cite{SU11}. In the present case $k$ can acquire two
values: 1/4 and 3/4; so we have two irreducible representations.
The even subspace ${\cal H}_{0}$ corresponds to the
representation with $k=1/4$, and the odd subspace ${\cal H}_{1}$
corresponds to the case $k=3/4$. The unitary squeezing operators
$S(\xi,\lambda)$ of equation (\ref{2.8}) form a reducible 
representation since they act on both irreducible sectors.
More specifically, they form the $k=1/4$ irreducible representation
when they act on ${\cal H}_{0}$ only and the $k=3/4$ irreducible 
representation when they act on ${\cal H}_{1}$ only. Related to this
is the fact that
\begin{equation}
[S(\xi,\lambda),\Pi_{0}] = [S(\xi,\lambda),\Pi_{1}] = 0 .   
\label{2.12}    \end{equation}

The parity-dependent squeezing operator is defined as
\begin{equation}
\begin{array}{c}\vspace{0.2cm}
U(\xi_{0},\lambda_{0};\xi_{1},\lambda_{1}) = 
S(\xi_{0},\lambda_{0}) \Pi_{0} + S(\xi_{1},\lambda_{1}) \Pi_{1}  \\
\xi_{j} = -r_{j} \exp(-i\theta_{j}) \;\;\;\;\;\;\;\; j = 0,1 .
\end{array}    \label{2.13}    
\end{equation}
This is a generalization of the ordinary squeezing operator 
(\ref{2.8}). Only in the special case
\begin{equation}
r_{0} = r_{1} \;\;\;\;\;\;\;\;\; 
\theta_{0} = \theta_{1} \;\;\;\;\;\;\;\;\;
\lambda_{0} = \lambda_{1}     \label{2.14}  
\end{equation}
does the operator (\ref{2.13}) reduce to the operator (\ref{2.8}).
The parity-dependent squeezing operator squeezes independently each 
irreducible sector. Acting with this operator on the Glauber
coherent state \cite{Gla}
\begin{equation}
|\beta\rangle = e^{-|\beta|^2/2} \sum_{n=0}^{\infty}
\frac{\beta^{n}}{\sqrt{n!}} |n\rangle       \label{2.15}
\end{equation}
we obtain the parity-dependent squeezed state:
\begin{equation}
|\beta;\xi_{0},\lambda_{0};\xi_{1},\lambda_{1}\rangle = 
U(\xi_{0},\lambda_{0};\xi_{1},\lambda_{1}) |\beta\rangle 
= S(\xi_{0},\lambda_{0}) \Pi_{0} |\beta\rangle + 
S(\xi_{1},\lambda_{1}) \Pi_{1} |\beta\rangle .  \label{2.16} 
\end{equation}
In the special case (\ref{2.14}), this state reduces to the 
ordinary squeezed state \cite{SS1}. Note that the states 
$\Pi_{0} |\beta\rangle$ and $\Pi_{1} |\beta\rangle$ 
(with a suitable normalization) are the even 
and odd coherent states \cite{EOCS}, which are special cases of 
macroscopic quantum superpositions also known as the 
Schr\"{o}dinger-cat states \cite{Sch_cat}. We see that
the parity-dependent squeezed states  
$|\beta;\xi_{0},\lambda_{0};\xi_{1},\lambda_{1}\rangle$
can be viewed as superpositions of two differently squeezed 
Schr\"{o}dinger-cat states.

The overlap of two parity-dependent squeezed states with the same 
squeezing parameters and different coherent amplitudes is
\begin{equation}
\langle \alpha;\xi_{0},\lambda_{0};\xi_{1},\lambda_{1}
|\beta;\xi_{0},\lambda_{0};\xi_{1},\lambda_{1}\rangle
= \langle \alpha|\beta \rangle 
= \exp\left(-\frac{1}{2}|\alpha|^2 -\frac{1}{2}|\beta|^2
+ \alpha^{\ast}\beta \right)     \label{2.17}      
\end{equation}
where we have used the unitarity of the parity-dependent squeezing 
operator. We also multiply the identity resolution \cite{Gla}
\begin{equation}
\frac{1}{\pi}\int d^{2}\!\beta\, |\beta\rangle\langle\beta| = 1 
\label{2.18}    \end{equation}
by the operator $U(\xi_{0},\lambda_{0};\xi_{1},\lambda_{1})$ on the
left and by its Hermitian conjugate 
$U^{\dagger}(\xi_{0},\lambda_{0};\xi_{1},\lambda_{1})$ on the right
in order to prove another form of the identity resolution:
\begin{equation}
\frac{1}{\pi}\int d^{2}\!\beta\, 
|\beta;\xi_{0},\lambda_{0};\xi_{1},\lambda_{1}\rangle 
\langle \beta;\xi_{0},\lambda_{0};\xi_{1},\lambda_{1}| = 1 .
\label{2.19}    \end{equation}
Equations (\ref{2.17}) and (\ref{2.19}) show that the set of the 
states $\{ |\beta;\xi_{0},\lambda_{0};\xi_{1},\lambda_{1}\rangle \}$
with fixed squeezing parameters and all the complex numbers $\beta$
forms an overcomplete basis in the Hilbert space ${\cal H}$.

\subsection{Parity-dependent Bogoliubov transformations}

\noindent
We introduce the parity-dependent Bogoliubov transformations:
\begin{equation}
\begin{array}{l}\vspace{0.2cm}
b \equiv U(\xi_{0},\lambda_{0};\xi_{1},\lambda_{1}) a
U^{\dagger}(\xi_{0},\lambda_{0};\xi_{1},\lambda_{1})  \\
b^{\dagger} \equiv U(\xi_{0},\lambda_{0};\xi_{1},\lambda_{1}) 
a^{\dagger} U^{\dagger}(\xi_{0},\lambda_{0};\xi_{1},\lambda_{1}) .  
\end{array}   \label{2.20}    
\end{equation}
The commutation relation is preserved for the parity-dependent 
Bogoliubov quasiparticles:
\begin{equation}
[b,b^{\dagger}] = [a,a^{\dagger}] = 1 .  \label{2.21}   
\end{equation}
In the special case (\ref{2.14}), the operators $b$ and 
$b^{\dagger}$ are linear combinations of $a$ and $a^{\dagger}$:
\begin{equation}
\begin{array}{l}\vspace{0.2cm}
b = \mu_{j} a + \nu_{j} a^{\dagger}  \\
b^{\dagger} = \nu_{j}^{\ast} a + \mu_{j}^{\ast} a^{\dagger} 
\end{array}   \label{2.22}    
\end{equation}
where $j$ is either 0 or 1, and we use the notation 
\begin{equation}
\begin{array}{l}\vspace{0.2cm}
\mu_{j} \equiv \cosh r_{j} \exp(-i\lambda_{j})  \\ \vspace{0.2cm}
\nu_{j} \equiv \sinh r_{j} \exp[-i(\theta_{j}+\lambda_{j})]  \\
|\mu_{j}|^{2} - |\nu_{j}|^{2} = 1 .
\end{array}   \label{2.23}    
\end{equation}
However, in general the transformation (\ref{2.20}) is much more
complicated and $b$ and $b^{\dagger}$ are not linear combinations of 
$a$ and $a^{\dagger}$. Using equations (\ref{2.20}), we easily prove:
\begin{equation}
U(\xi_{0},\lambda_{0};\xi_{1},\lambda_{1}) f(a,a^{\dagger})
U^{\dagger}(\xi_{0},\lambda_{0};\xi_{1},\lambda_{1})
= f(b,b^{\dagger}) .   \label{2.24} 
\end{equation}
It is easily seen that the parity-dependent squeezed states of 
equation (\ref{2.16}) are the
ordinary coherent states with respect to the Bogoliubov
quasiparticles. For example, they are eigenstates of $b$:
\begin{equation}
b |\beta;\xi_{0},\lambda_{0};\xi_{1},\lambda_{1}\rangle
= \beta |\beta;\xi_{0},\lambda_{0};\xi_{1},\lambda_{1}\rangle .
\label{2.25}     \end{equation}
We also can introduce the ``$b$-position'' operator
\begin{equation}
x_{b} = U(\xi_{0},\lambda_{0};\xi_{1},\lambda_{1}) x
U^{\dagger}(\xi_{0},\lambda_{0};\xi_{1},\lambda_{1}) =
\frac{b+b^{\dagger}}{\sqrt{2}}       \label{2.26}
\end{equation}
whose eigenstates are the ``$b$-position'' states:
\begin{equation}
\begin{array}{l}\vspace{0.2cm}
|x\rangle_{b} = U(\xi_{0},\lambda_{0};\xi_{1},\lambda_{1}) 
|x\rangle  
\\ x_{b}|x\rangle_{b} = x|x\rangle_{b} .
\end{array}    \label{2.27}
\end{equation}
The overlap of the parity-dependent squeezed states with the 
``$b$-position'' states is a simple Gaussian:
\begin{equation}
{}_{b}\!\langle x|\beta;\xi_{0},\lambda_{0};
\xi_{1},\lambda_{1}\rangle = \langle x|\beta\rangle  
= \pi^{-1/4} \exp\left[ -\frac{\beta}{2}(\beta^{\ast}-\beta)
-\left(\beta-\frac{x}{\sqrt{2}}\right)^{2} \right]  .      
\label{2.28}     \end{equation}
Consequently, the variances of the ``$b$-position'' and 
``$b$-momentum'' over the parity-dependent squeezed states are
\begin{equation}
(\Delta x_{b})^{2} = (\Delta p_{b})^{2} = 1/2 . \label{2.29}
\end{equation}
The uncertainties with respect to the ordinary position and momentum 
are discussed in section \ref{sec:3b}.

\subsection{The parity-dependent Hamiltonian}

\noindent
Here we give a Hamiltonian that can produce the parity-dependent 
squeezed states. Using the identity 
\begin{equation}
\exp(A\Pi_{0}+B\Pi_{1}) = \Pi_{0}\exp(A) + \Pi_{1}\exp(B) 
\label{2.30}    \end{equation}
(where $A$ and $B$ are operators that commute with $\Pi_{0}$ and 
$\Pi_{1}$), we easily see that quantum systems governed by the 
Hamiltonian
\begin{equation}
H = \omega a^{\dagger}a + \Pi_{0}(g_{0} a^{\dagger 2} +
g_{0}^{\ast} a^{2}) + \Pi_{1}(g_{1} a^{\dagger 2} +
g_{1}^{\ast} a^{2})     \label{2.31} 
\end{equation}
will evolve ordinary coherent states $|\beta\rangle$ into the
parity-dependent squeezed states of equation (\ref{2.16}). From the 
relations
$\Pi_{0} = (1+P)/2$ and $\Pi_{1} = (1-P)/2$, we see that the
Hamiltonian (\ref{2.31}) contains the parity operator
$P=\exp(i\pi a^{\dagger}a)$. In the special case $g_{0}=g_{1}$,
the Hamiltonian (\ref{2.31}) reduces to the Hamiltonian
\begin{equation}
H = \omega a^{\dagger}a + g a^{\dagger 2} + g^{\ast} a^{2}
\label{2.32}     \end{equation}
that describes the degenerate down-conversion process in
which the usual single-mode squeezed states are produced.

\section{Quantum statistical properties}
\setcounter{equation}{0}

\subsection{Photon statistics}

\noindent
The number-state decomposition of the ordinary squeezed states
$|\beta;\xi,\lambda\rangle = S(\xi,\lambda)|\beta\rangle$
is given by \cite{SS1}
\begin{equation}
|\beta;\xi,\lambda\rangle = \frac{1}{\sqrt{\mu}} \exp\left(
-\frac{|\beta|^{2}}{2}+\frac{\nu^{\ast}}{2\mu}\beta^{2} \right) 
\sum_{n=0}^{\infty} \frac{1}{\sqrt{n!}} \left(
\frac{\nu}{2\mu} \right)^{n/2}\! H_{n}\left(\frac{\beta
}{\sqrt{2\mu\nu}} \right) |n\rangle     \label{3.1} 
\end{equation}
where the parameters $\mu$ and $\nu$ are defined
according to equation (\ref{2.23}), and $H_{n}(z)$ are the Hermite
polynomials. By using equation (\ref{3.1}), we easily obtain the
number-state decomposition of the parity-dependent squeezed states:
\begin{eqnarray}
|\beta;\xi_{0},\lambda_{0};\xi_{1},\lambda_{1}\rangle
& = & \sum_{j=0}^{1} \frac{1}{\sqrt{\mu_{j}}} \exp\left(
-\frac{|\beta|^{2}}{2}+\frac{\nu_{j}^{\ast}}{2\mu_{j}}\beta^{2} 
\right)
\sum_{n=0}^{\infty} \frac{1}{\sqrt{(2n+j)!}} \left(
\frac{\nu_{j}}{2\mu_{j}} \right)^{(2n+j)/2} \nonumber \\  
& & \times H_{2n+j}\left(\frac{\beta
}{\sqrt{2\mu_{j}\nu_{j}}} \right) |2n+j\rangle .    \label{3.2} 
\end{eqnarray}
Then we find that the photon-number distribution
$P(n) = |\langle n|\beta;\xi_{0},\lambda_{0};\xi_{1},\lambda_{1}
\rangle|^{2}$ is given by
\begin{equation}
P(n) = \exp(-|\beta|^{2} + |\beta|^{2}\tanh r_{j} \cos 2\psi_{j})
\frac{ \tanh^{n}\! r_{j} }{ 2^{n} n! \cosh r_{j} } 
\left| H_{n}\left( \frac{ |\beta |e^{i\psi_{j}} }{ 
\sqrt{\sinh 2r_{j}} } \right) \right|^{2}   \label{3.3}     
\end{equation}
where the index $j$ is 0 for even $n$ and 1 for odd $n$. We use the
notation
\begin{eqnarray}
& & \psi_{j} \equiv \phi_{\beta} + \lambda_{j}
+ \frac{1}{2} \theta_{j}       \label{3.4}   \\
& & \phi_{\beta} \equiv {\rm arg}\, \beta .  \label{3.5}
\end{eqnarray}
Numerical results are shown in figures 1 and 2. The known oscillations 
\cite{oscil} in the photon-number distribution of the ordinary 
squeezed states appear also in the parity-dependent case. Due to the
fact that the distributions for even and odd photon numbers depend
on different parameters, these oscillations can be enhanced or 
decreased by suitable choice of the parameters. The distribution
$P(n)$ of equation (\ref{3.3}) shows oscillations of two types: ``slow''
oscillations that follow from smooth oscillations of the Hermite
polynomials, and ``rapid'' oscillations (or, more precisely, sharp
jumps) between even and odd values of $n$. These jumps follow from
the fact that the Hermite polynomials for even and odd values of $n$
behave in a different manner. These features exist also for the
ordinary squeezed states, but in the parity-dependent case it is
possible to change independently the behavior of the even and odd
parts.

The characteristic function
\begin{equation}
F(z) = \sum_{n=0}^{\infty} z^{n} P(n)    \label{3.6}
\end{equation}
allows us to calculate all the normally ordered moments of the 
number operator $N = a^{\dagger}a$:
\begin{equation}
\langle a^{\dagger p} a^{p} \rangle = \langle N(N-1)\cdots 
(N-p+1) \rangle = \left. \frac{ \partial^{p} F }{ \partial z^{p} }
\right|_{z=1} .      \label{3.7}
\end{equation}
Combining equations (\ref{3.3}) and (\ref{3.6}) and using summation
theorems for the Hermite polynomials \cite{Erd}, we get
\begin{equation}
F(z) = \frac{e^{-|\beta|^{2}}}{2} \sum_{j=0}^{1} \tau_{j}
\left[ e^{z\tau_{j}^{2}|\beta|^{2}} + (-1)^{j} 
e^{-z\tau_{j}^{2}|\beta|^{2}} \right] \exp\left[ |\beta|^{2}
(1-\tau_{j}^{2}z^{2}) \tanh r_{j} \cos 2\psi_{j} \right] 
\label{3.8}      \end{equation}
where we have defined
\begin{equation}
\tau_{j} \equiv \left( \cosh^{2}\! r_{j} - z^{2}
\sinh^{2}\! r_{j} \right)^{-1/2} .  \label{3.9} 
\end{equation}
Using equations (\ref{3.7}) and (\ref{3.8}), we find analytic
expressions for the first and the second moments:
\begin{eqnarray}
& & \langle a^{\dagger} a \rangle = \frac{1}{2} \sum_{j=0}^{1}
\left[ A_{j}^{(+)} + (-1)^{j} e^{-2|\beta|^{2}} A_{j}^{(-)}
\right]   \label{3.10} \\
& & \langle a^{\dagger 2} a^{2} \rangle = \frac{1}{2} \sum_{j=0}^{1}
\left\{ \left[ \left( A_{j}^{(+)} \right)^{2} + B_{j}^{(+)}
\right] + (-1)^{j} e^{-2|\beta|^{2}}  \left[ \left( A_{j}^{(-)} 
\right)^{2} + B_{j}^{(-)} \right] \right\}   \label{3.11}
\end{eqnarray}
where we have defined
\begin{eqnarray}
A_{j}^{(\pm)} & \equiv & \sinh^{2}\! r_{j} \pm
|\beta|^{2} \cosh 2r_{j} - |\beta|^{2} \sinh 2r_{j} 
\cos 2\psi_{j}  \label{3.12} \\
B_{j}^{(\pm)} & \equiv & \sinh^{2}\! r_{j} \cosh 2r_{j}
\pm 2|\beta|^{2} \sinh^{2}\! r_{j} (1 + 2\cosh 2r_{j}) \nonumber \\
& & - |\beta|^{2} \sinh 2r_{j}(1 + 4\sinh^{2}\! r_{j}) 
\cos 2\psi_{j}  .            \label{3.13}
\end{eqnarray}
The second-order correlation function 
\begin{equation}
g^{(2)} = \frac{ \langle a^{\dagger 2} a^{2} \rangle }{
\langle a^{\dagger} a \rangle^{2} } = \frac{ \langle N^{2}
\rangle - \langle N \rangle }{ \langle N \rangle^{2} }
  \label{3.14}
\end{equation}
can be calculated from equations (\ref{3.10}) and (\ref{3.11}). 

Numerical calculations show that in the case $r_{0}=0$ (only the 
odd component is squeezed) antibunching is relatively weak. In 
this case the minimum value of $g^{(2)}$ is
approximately 0.75 for $r_{1} = 0.3$, $\psi_{1} = 0$, $|\beta|
\simeq 1$. Much more strong antibunching is obtained for
$r_{1}=0$ (only the even component is squeezed). This situation is
shown in figure 3. We see that the parity-dependent squeezed states 
are antibunched for small
values of $r_{0}$ and $|\beta|$. Also, we find that maximum
antibunching is achieved for $\psi_{0} = 0$, i.e., for squeezing
in the direction of the displacement of the initial coherent state.
The dependence of $g^{(2)}$ on $\psi_{0}$ is seen from figure 4.
When $\psi_{0} = 0$, $r_{1}=0$, very strong antibunching can
be achieved for very small values of $r_{0}$ and $|\beta|$. This
is shown in figure 5 where $g^{(2)}$ is presented for the 
parity-dependent and ordinary squeezed states. We see that the 
parity-dependent squeezed states with $r_{1}=0$, $\psi_{0} = 0$ 
exhibit stronger antibunching
than the ordinary squeezed states with $r=r_{0}$, $\psi=0$.

\subsection{Position and momentum uncertainties}
\label{sec:3b}

\noindent
The position and momentum operators are defined as
\begin{eqnarray} 
& & x = \frac{1}{\sqrt{2}} (a^{\dagger}+a)  \label{3.15}  \\
& & p = \frac{i}{\sqrt{2}} (a^{\dagger}-a) . \label{3.16}
\end{eqnarray}
Their variances
\begin{eqnarray} 
& & (\Delta x)^{2} = \langle x^{2} \rangle - \langle x \rangle^{2} 
\label{3.17}  \\
& & (\Delta p)^{2} = \langle p^{2} \rangle - \langle p \rangle^{2} 
\label{3.18}   \end{eqnarray}
obey the Heisenberg uncertainty relation
\begin{equation}
(\Delta x)^{2} (\Delta p)^{2} \geq \frac{1}{4} .  \label{3.19}
\end{equation}
The $x$-representation of the parity-dependent squeezed states can 
be calculated to be
\begin{eqnarray} 
\Psi(x) & = & \langle x
|\beta;\xi_{0},\lambda_{0};\xi_{1},\lambda_{1}\rangle =
\frac{1}{2} \left( \frac{1}{\pi} \right)^{1/4} \exp(-|\beta|^{2}/2)
\sum_{j=0}^{1} \frac{1}{
\sqrt{\mu_{j}-\nu_{j}} } \exp\left( - \frac{
\mu_{j}^{\ast}-\nu_{j}^{\ast}}{\mu_{j}-\nu_{j}} 
\frac{\beta^{2}}{2} \right) \nonumber \\ & & 
\times \left[ \exp\left( \frac{\sqrt{2}\beta x}{
\mu_{j}-\nu_{j}} \right) + (-1)^{j}
\exp\left( -\frac{\sqrt{2}\beta x}{\mu_{j}-\nu_{j}} \right) 
\right] \exp\left(- \frac{\mu_{j}+\nu_{j}}{\mu_{j}-\nu_{j}} 
\frac{x^{2}}{2} \right) .  \label{3.20}
\end{eqnarray}
Moments of the position operator are found by evaluating
Gaussian integrals:
\begin{eqnarray} 
\langle x \rangle & = & \frac{ e^{-|\beta|^{2}} }{ 2\sqrt{2} } 
\sum_{j,l=0}^{1} (1 - \delta_{jl} )\, \Omega_{jl}^{3/2} 
\left[ V_{jl}^{(+)} e^{\Omega_{jl}|\beta|^{2}}
+ (-1)^{j} V_{jl}^{(-)} e^{-\Omega_{jl}|\beta|^{2}} \right]
\nonumber \\ & & \times \exp\left\{ i\Omega_{jl}{\rm Im}\,
[\beta^{\ast 2}(\mu_{j}\nu_{l}-\mu_{l}\nu_{j})] \right\} \\
\label{3.21}
\langle x^{2} \rangle & = & \frac{1}{4} \sum_{j=0}^{1}\left\{\left(
|\mu_{j}-\nu_{j}|^{2} + \left[ V_{jj}^{(+)} \right]^{2} \right)
+ (-1)^{j} e^{-2|\beta|^{2}}
\left( |\mu_{j}-\nu_{j}|^{2} + \left[ V_{jj}^{(-)} \right]^{2} 
\right) \right\}   \label{3.22}
\end{eqnarray}
where $\delta_{jl}$ is the Kronecker symbol, and we have defined
\begin{eqnarray} 
& & \Omega_{jl} \equiv (\mu_{j}\mu_{l}^{\ast} -
\nu_{j}\nu_{l}^{\ast})^{-1}   \label{3.23} \\
& & V_{jl}^{(\pm)} \equiv \beta^{\ast}(\mu_{j}-\nu_{j}) 
\pm \beta(\mu_{l}^{\ast}-\nu_{l}^{\ast}) .  \label{3.24}
\end{eqnarray}
The $p$-representation of the parity-dependent 
squeezed states and moments of the momentum operator can be 
obtained analogously. Then we also can calculate
the uncertainty product $(\Delta x)^{2} (\Delta p)^{2}$.

We have studied numerically the behavior of the variance 
$(\Delta x)^{2}$ for $\phi_{\beta}=\lambda_{0}=\lambda_{1}=0$.
Calculations show that the parity-dependent squeezed states are 
squeezed in the $x$-direction
for $\theta_{0}=\theta_{1}=0$. Squeezing in the $p$-direction
is obtained, accordingly, for $\theta_{0}=\theta_{1}=\pm\pi$.
In the case $r_{1}=0$, $\theta_{0}=0$ we find that
$(\Delta x)^{2} \rightarrow \frac{1}{2}\exp(-2r_{0})$ as
$|\beta| \rightarrow 0$. As $|\beta|$ increases, $(\Delta x)^{2}$
increases too. In the case $r_{0}=0$, $\theta_{1}=0$ the situation
is essentially different, as shown in figure 6. For small values of
$|\beta|$ ($|\beta| \ll 1$), $(\Delta x)^{2}$ approaches the
coherent-state value 1/2. As $|\beta|$ increases, $(\Delta x)^{2}$
at first decreases below 1/2 (squeezing), reaches a minimum
and then increases monotonically.
The uncertainty product $(\Delta x)^{2} (\Delta p)^{2}$ is plotted
in figure 7 as a function of $|\beta|$ for the case $\phi_{\beta}=0$,
$r_{1}=0$, $\theta_{0}=\lambda_{0}=\theta_{1}=\lambda_{1}=0$ and
various values of $r_{0}$. The uncertainty product is always
greater than its minimum allowed value 1/4. This value is achieved
only in the limit $|\beta| \rightarrow 0$. Recall that the 
uncertainty product in the variables $x_{b}$, $p_{b}$ is always 1/4.

\section{$Q$ and Wigner functions}
\setcounter{equation}{0}

\noindent
Useful information about the field state can be inferred from
phase-space quasiprobability distributions. We start
from the $Q(\alpha)$ function:
\begin{equation}
Q(\alpha) = \frac{1}{\pi} | \langle \alpha
|\beta;\xi_{0},\lambda_{0};\xi_{1},\lambda_{1}\rangle |^{2} .
\label{4.1}   \end{equation}
A strightforward calculation gives
\begin{eqnarray} 
\langle\alpha|\beta;\xi_{0},\lambda_{0};\xi_{1},\lambda_{1}\rangle
& = & \frac{1}{2} e^{-(|\alpha|^{2}+|\beta|^{2})/2}
\sum_{j=0}^{1} \frac{1}{\sqrt{\mu_{j}}} 
\left[ e^{\alpha^{\ast}\beta/\mu_{j}} + (-1)^{j}
e^{-\alpha^{\ast}\beta/\mu_{j}} \right]  \nonumber \\
& & \times \exp\left( \frac{\nu_{j}^{\ast}}{2\mu_{j}} \beta^{2} 
- \frac{\nu_{j}}{2\mu_{j}} \alpha^{\ast 2} \right) .  \label{4.2} 
\end{eqnarray}
The Wigner function is given by \cite{Wig}
\begin{equation}
W(x,p) = \frac{1}{\pi} \int_{-\infty}^{\infty} \Psi(x+s)
\Psi^{\ast}(x-s)\, e^{-2ips} ds .    \label{4.3}
\end{equation}
Using equation (\ref{3.20}) for the $x$-representation wave function
and evaluating the Gaussian integrals, we find the 
following result:
\begin{eqnarray} 
W(x,p) & = & \frac{1}{4\pi} \sum_{j,l=0}^{1}
\Omega_{jl}^{1/2} \exp\left(-|\beta|^{2} - Z_{jl} - T_{jl}
\frac{x^{2}}{2} \right) \nonumber \\
& & \times \left\{ (-1)^{j} \exp\left[ \frac{ \left(R_{jl}x
-2ip - K_{jl} \right)^{2} }{ 2T_{jl} } + L_{jl}x \right] \right.
\nonumber \\ & & +(-1)^{l} \exp\left[ \frac{ \left(
R_{jl}x-2ip + K_{jl} \right)^{2} }{ 2T_{jl} } - L_{jl}x \right]
\nonumber \\ & & +(-1)^{j+l} \exp\left[ \frac{ \left(
R_{jl}x-2ip + L_{jl} \right)^{2} }{ 2T_{jl} } - K_{jl}x \right]
\nonumber \\ & & \left. + \exp\left[ \frac{ \left(R_{jl}x
-2ip - L_{jl} \right)^{2} }{ 2T_{jl} } + K_{jl}x \right]
\right\}    \label{4.4}
\end{eqnarray}
where we have defined
\begin{eqnarray} 
& & Z_{jl} \equiv  \frac{
\mu_{j}^{\ast}-\nu_{j}^{\ast}}{\mu_{j}-\nu_{j}} 
\frac{\beta^{2}}{2} + \frac{\mu_{l}-\nu_{l}}{
\mu_{l}^{\ast}-\nu_{l}^{\ast}}
\frac{\beta^{\ast 2}}{2}  \\
& & T_{jl} \equiv  \frac{\mu_{j}+\nu_{j}}{\mu_{j}-\nu_{j}} 
+ \frac{\mu_{l}^{\ast}+\nu_{l}^{\ast}}{
\mu_{l}^{\ast}-\nu_{l}^{\ast}}   \\
& & R_{jl} \equiv - \frac{\mu_{j}+\nu_{j}}{\mu_{j}-\nu_{j}} 
+ \frac{\mu_{l}^{\ast}+\nu_{l}^{\ast}}{
\mu_{l}^{\ast}-\nu_{l}^{\ast}}   \\
& & K_{jl} \equiv  \frac{\sqrt{2}\beta}{\mu_{j}-\nu_{j}} 
+ \frac{\sqrt{2}\beta^{\ast}}{
\mu_{l}^{\ast}-\nu_{l}^{\ast}}   \\
& & L_{jl} \equiv - \frac{\sqrt{2}\beta}{\mu_{j}-\nu_{j}} 
+ \frac{\sqrt{2}\beta^{\ast}}{
\mu_{l}^{\ast}-\nu_{l}^{\ast}}  
\end{eqnarray}
and $\Omega_{jl}$ is defined by equation (\ref{3.23}). 
We see that in the parity-dependent case both the $Q$ and Wigner 
functions are given by a superposition of a number of Gaussians 
and not by a single Gaussian as in the ordinary case. 

For the parity-dependent squeezing the interference in phase 
space produces $Q$ and Wigner functions of interesting forms.
Figure 8 shows that the $Q(\alpha)$ function for the case of 
strongly squeezed even component is similar to that of a number
eigenstate. The $Q(\alpha)$ functions shown in figures 9 and 10 
have $r_{0}=r_{1}$, and $\lambda_{0}=\lambda_{1}$, but 
$\theta_{0} \neq \theta_{1}$. And yet this is enough to split the
Gaussian $Q$ function of an ordinary squeezed state into three
Gaussians in figure 9 and five Gaussians in figure 10.
Some examples of the Wigner function for the parity-dependent 
squeezed states are shown in figures 11 and 12. In figure 11 we see 
a big peak along the line $p=0$ and a smaller ``wave'' along the 
line $x=0$. When these two structures intersect near the origin two 
sharp negative peaks are produced. Besides, two high positive peaks 
become at the intersection of the big peak with two smaller 	
Gaussians perpendicular to it. In figure 12 a very impressive 
interference occurs along the line $x=0$ where sharp positve and 
negative peaks alternate.

\section{Conclusions}
\setcounter{equation}{0}

\noindent
In this paper we have introduced the concept of parity-dependent
squeezing for the single-mode light field. It is based on the fact
that squeezing transformations are elements of the SU(1,1) Lie
group, and therefore we proposed a squeezing operator that acts
differently on distinct irreducible representations of SU(1,1). 
For the case of single-mode squeezing this operator is parity
dependent. We have considered the
parity-dependent Bogoliubov transformations and parity-dependent
Bogoliubov quasiparticles. A parity-dependent quadratic
Hamiltonian has been given that evolves coherent states into the
parity-dependent squeezed states. Quantum statistical properties
of these states have been studied in detail. We have found 
interesting nonclassical features such as strong oscillations in the
photon-number distribution, strong antibunching and quadrature
squeezing. Results for the $Q$ and Wigner functions show that 
parity-dependent squeezing considered in this paper leads to very
interesting interference effects in phase space which are absent in 
ordinary squeezing.

\section*{Acknowledgments}

\noindent
CB gratefully acknowledges the financial help from the Technion.
AM was supported by the Fund for Promotion of Research
at the Technion, by the Technion -- VPR Fund, and by the Harry
Werksman Research Fund.
AV gratefully acknowledges support from the British council in 
the form of a travel grant.

\begin{flushleft}

\end{flushleft}

\section*{Figure captions}
%1
\begin{description}
\item{Figure 1:}
The photon-number distribution $P(n)$ for a 
parity-dependent squeezed state with $|\beta|=4$, $r_{0}=0.5$, 
$\psi_{0}=\pi/2$, $r_{1}=0.1$, $\psi_{1}=\pi/2$. The wide peaks 
on the left and right sides contain sharp peaks for odd and even 
values of $n$, respectively.
\item{Figure 2:}
The photon-number distribution $P(n)$ for a 
parity-dependent squeezed state with $|\beta|=4$, $r_{0}=0.5$, 
$\psi_{0}=0$, $r_{1}=0.1$, $\psi_{1}=\pi/2$. The wide peaks on 
the left and right sides contain sharp peaks for even and odd 
values of $n$, respectively.
\item{Figure 3:}
The second-order correlation function $g^{(2)}$ versus
$|\beta|$ for $r_{1}=0$, $\psi_{0}=\psi_{1}=0$ and various values
of $r_{0}$. Antibunching appear for $r_{0} < 0.48$, and the smaller
$r_{0}$, the stronger antibunching.
\item{Figure 4:}
The second-order correlation function $g^{(2)}$ versus
$|\beta|$ for $r_{0}=0.05$, $r_{1}=0$, $\psi_{1}=0$ and 
various values of $\psi_{0}$. Antibunching is strong for 
$\psi_{0}$ near zero, but rapidly weakens as $\psi_{0}$
increases.
\item{Figure 5:}
The second-order correlation function $g^{(2)}$ versus
$|\beta|$ for parity-dependent squeezed states with $r_{1}=0$, 
$\psi_{0}=0$ (solid line) and ordinary squeezed states with 
$r=r_{0}$, $\psi=0$ (dashed line). Various values of $r=r_{0}$
are considered.
\item{Figure 6:}
The variance $(\Delta x)^{2}$ as a function of $|\beta|$
for $\phi_{\beta}= 0$, $r_{0}=0$, $\theta_{0}=\lambda_{0}=\theta_{1}
=\lambda_{1}=0$ and various values of $r_{1}$. For moderate values
of $|\beta|$ and $r_{1}$ the state is squeezed in the $x$-direction,
$(\Delta x)^{2} < 1/2$.
\item{Figure 7:}
The uncertainty product $(\Delta x)^{2}(\Delta p)^{2}$
as a function of $|\beta|$ for $\phi_{\beta}= 0$, $r_{1}=0$, 
$\theta_{0}=\lambda_{0}=\theta_{1}=\lambda_{1}=0$ and various values
of $r_{0}$. The dashed line is the minimum available value 1/4.
\item{Figure 8:}
The function $Q(\alpha)$ for a parity-dependent squeezed 
state with $|\beta|=1$, $\phi_{\beta}=0$, $r_{0}=4$, $r_{1}=0$,
$\theta_{0}=\theta_{1}=0$, $\lambda_{0}=\lambda_{1}=0$.
\item{Figure 9:}
The function $Q(\alpha)$ for a parity-dependent squeezed 
state with $|\beta|=3$, $\phi_{\beta}=0$, $r_{0}=r_{1}=3$,
$\theta_{0}=0$, $\theta_{1}=\pi$, $\lambda_{0}=\lambda_{1}=0$.
\item{Figure 10:}
The function $Q(\alpha)$ for a parity-dependent squeezed 
state with $|\beta|=5$, $\phi_{\beta}=0$, $r_{0}=r_{1}=3$,
$\theta_{0}=0$, $\theta_{1}=\pi$, $\lambda_{0}=\lambda_{1}=0$.
\item{Figure 11:}
The Wigner function for a parity-dependent squeezed 
state with $|\beta|=3$, $\phi_{\beta}=0$, $r_{0}=3$, $r_{2}=0$,
$\theta_{0}=\pi$, $\theta_{1}=0$, $\lambda_{0}=\lambda_{1}=0$.
\item{Figure 12:}
The Wigner function for a parity-dependent squeezed 
state with $|\beta|=8$, $\phi_{\beta}=0$, $r_{0}=r_{1}=3$,
$\theta_{0}=0$, $\theta_{1}=\pi$, $\lambda_{0}=\lambda_{1}=0$.

\end{description}


\begin{thebibliography}{99}
%\vspace{0.2cm}
%
\bibitem{SS1} Stoler D 1970 {\em Phys. Rev.} D {\bf 1} 3217; 
1971 {\em Phys. Rev.} D {\bf 4} 2308 (1971) \linebreak
Yuen H P 1976 {\em Phys. Rev.} A {\bf 13} 2226 \linebreak
Hollenhorst J N 1979 {\em Phys. Rev.} D {\bf 19} 1669 \linebreak
Walls D F 1983 {\em Nature} {\bf 306} 141 
%
\bibitem{SS2} Caves C M and Schumaker B L 1985 {\em Phys. Rev.} A 
{\bf 31} 3068  \linebreak
Schumaker B L and Caves C M 1985 {\em Phys Rev.} A {\bf 31} 3093  
\linebreak
Schumaker B L 1986 {\em Phys. Rep.} {\bf 135} 317 
%
\bibitem{SS3} Loudon R and Knight P L 1987 {\em J. Mod. Opt.}   
{\bf 34} 709 \linebreak
Teich M C and Saleh B E A 1990 {\em Quantum Opt.} {\bf 1} 153 \linebreak
Fabre C 1992 {\em Phys. Rep.} {\bf 219} 215 
%
\bibitem{SMD} Simon R, Mukunda N and Dutta B 1994 {\em Phys. Rev.} A
{\bf 49} 1567 and references therein
%
\bibitem{oscil} Schleich W and Wheeler J A 1987 {\em Nature} 
{\bf 326} 574 \linebreak
Vourdas A and Weiner R 1987 {\em Phys. Rev.} A {\bf 36} 5866 
%
\bibitem{Vou} Vourdas A 1992 {\em Phys. Rev.} A {\bf 46} 442
%
\bibitem{SU11} Bargmann V 1947 {\em Ann. Math.} {\bf 48} 568 
%
\bibitem{Gla} Glauber R J 1963 {\em Phys. Rev.} {\bf 130} 2529; 
1963 {\em Phys. Rev.} {\bf 131} 2766
%
\bibitem{EOCS} Dodonov V V, Malkin I A and Man'ko V I 1974
{\em Physica} {\bf 72} 597 
%
\bibitem{Sch_cat} Schr\"{o}dinger E 1935 {\em Naturwissenschaften} 
{\bf 23} 844 
%
\bibitem{Erd} Erd\'{e}lyi {\em et al} (eds) 1953
{\em Bateman Manuscript Project: Higher Transcendental Functions}
(New York: McGraw-Hill) vol 2, ch X
%
\bibitem{Wig} Wigner E P 1932 {\em Phys. Rev.} {\bf 40} 749 \linebreak
Hillery M, O'Connell R F, Scully M O and Wigner E P 1984
{\em Phys. Rep.} {\bf 106} 121 
%
\end{thebibliography}
\end{document}